\documentclass[aps,prl,twocolumn,groupedaddress,showpacs]{revtex4}
\usepackage{graphics}
\usepackage{color}
\usepackage{amsmath,amsbsy,bm,amssymb}

\usepackage[]{graphicx}
\DeclareGraphicsExtensions{.pdf}

\begin{document}

\title{Beyond the French Flag Model:  Exploiting Spatial and Gene Regulatory Interactions for Positional Information}

\author{Patrick Hillenbrand${}^a$, Ulrich Gerland${}^a$, Ga\v{s}per Tka\v{c}ik${}^b$}
\affiliation{$^a$Physics of Complex Biosystems, Physics Department,Technical University of Munich, James-Franck-Str. 1, D-85748 Garching, Germany\\
$^b$Institute of Science and Technology Austria,
Am Campus 1, A-3400 Klosterneuburg, Austria}

\date{\today}

\begin{abstract}
A crucial step in the early development of multicellular organisms involves the establishment of spatial patterns of gene expression which later direct proliferating cells to take on different cell fates. These patterns enable the cells to infer their global position within a tissue or an organism by reading out local gene expression levels. The patterning system is thus said to encode \emph{positional information}, a concept that was formalized recently in the framework of information theory. Here we introduce a toy model of patterning in one spatial dimension, which can be seen as an extension of Wolpert's paradigmatic ``French Flag'' model, to patterning by several interacting, spatially coupled genes subject to intrinsic and extrinsic noise. Our model, a variant of an Ising spin system, allows us to systematically explore expression patterns that optimally encode positional information. We find that optimal patterning systems use positional cues, as in the French Flag model, together with gene-gene interactions to generate combinatorial codes for position which we call ``Counter'' patterns. Counter patterns can also be stabilized against noise and variations in system size or morphogen dosage by longer-range spatial interactions of the type invoked in the Turing model.  The simple setup proposed here qualitatively captures many of the experimentally observed properties of biological patterning systems and allows them to be studied in a single, theoretically consistent framework. 
\end{abstract}

\maketitle
\section{Introduction}
Shape and size are global properties of organisms and of their constituent parts. Yet organisms develop and grow by  processes that are intrinsically local: cell division, fate commitment and differentiation, migration, and death. To coordinate these processes appropriately, cells must reproducibly activate different gene expression programs in a manner that is positionally specified, or \emph{patterned}, within a tissue or a whole organism. Here we are interested in  essential conditions for pattern-forming systems to  support rich and robust positional specification of cells. Rather than focusing on any particular  organism, we analyze a minimal tractable model that can qualitatively reproduce pattering features across a diverse range of biological examples. In the process, we  illustrate  and formalize a number of concepts pertaining to developmental pattern formation.

One possibility for cells to acquire their position-dependent fates is to establish a field of developmental cues that the cells can ``read out'' to learn about their individual locations in an organism and hence to appropriately coordinate their behaviors. Often, these cues are gradients of patterning molecules, called morphogens. A morphogen gradient is a chemical ``coordinate system'' in which the organism's body plan is drawn \cite{driever1988}; in this analogy, reproducibility of developmental outcomes  is limited by the reliability with which physical locations map into morphogen concentrations \cite{Gregor:2007wz}. A canonical embodiment of this idea is the French Flag model, where a smooth spatial morphogen gradient activates downstream cell-fate-determining genes at different thresholds, creating stripes in a previously unpatterned tissue \cite{wolpert1969, dahmann2011}.

Patterning strategies need not rely on the existence of signals distributed throughout the organism. Organism or tissue boundaries are intrinsically different from the bulk, and one can envision local biophysical mechanisms that propagate information from the boundary into the bulk to set up a global pattern. The Turing model is an example of this kind, where local rules are expressed as a set of reaction-diffusion equations and the steady-state pattern is determined by the boundary and initial conditions \cite{turing1952}. A Turing mechanism in concert with a globally acting morphogen gradient has, for instance,  been found to control vertebrate digit formation~\cite{drossopoulou2000,tickle2006,sheth2012,raspopovic2014}. 

Changing focus from systems continuous in space, time, and concentration to discrete setups, one can think of cellular-automata-like models that, starting with a defined initial state at the boundary of a finite domain, propagate that state into the bulk, creating a discrete pattern in a lattice of cells. Similarity to such a local, rule-based mechanism can be found, for instance, in developmental notch signaling \cite{bray2006,hamada2014}. 

Ultimately, patterning dynamics, be it continuous or discrete, need not even lead to a well-defined steady state;  local transients (as opposed to instantaneous or average values) of patterning cues could mark the position, much like during signaling leading up to the  aggregation of \emph{Dictyostelium} cells \cite{gregor2010}, or in the clock-and-wavefront model for the generation of somites \cite{cooke1976,palmeirim1997,oates2012}.

While such models typically represent a gross simplification of reality, they do capture a fundamental property of biological pattern formation: local patterning cues, either in stationary state or during  a readout period, carry information about position relative to a global reference frame \cite{jaeger2004}. This property was introduced as ``positional information'' by Wolpert in his landmark paper almost fifty years ago \cite{wolpert1969}. Despite intense study in the last decades~\cite{wolpert1989,crauk2005,jaeger2006,Gregor:2007wz,kerszberg2007,jaeger2008,bollenbach2008,jaeger2009,he2010}, it has been difficult to come up with a formal definition of positional information and a corresponding measure that would quantify the regulatory power of a patterning system by, in essence, counting the maximal number of distinct cell fates that the system can reliably specify, irrespective of mechanistic detail. This is because the mapping between position and local cue values can be noisy or even ambiguous, and can be established by a diverse range of biophysical mechanisms.  Additionally, it would be attractive to build such a quantity on a strong theoretical foundation on one hand, while on the other ensure that it could be computed in various models of patterning or be tractably estimated from data. 

A candidate formalization of ``positional information'' that satisfies the above criteria, based on application of information theoretic ideas, has recently been proposed \cite{dubuis2013,tkacik2015}. Positional information can be seen as a generic measure of correlation, i.e., a mutual information~\cite{cover2012}, between position and local patterning cue values (e.g., morphogen expression levels). It is also closely related to information transmission through genetic regulatory networks that has been a subject of recent theoretical~\cite{Ziv:2007bo,Tkacik:2008us,tostevin2009,bowsher2014} and data-driven investigations~\cite{Tkacik:2008dq,cheong2011,selimkhanov2014,hansen2015}. Looking at anterior-posterior patterning in early \emph{Drosophila} embryo, the four primary gap genes were estimated to carry $4.2\pm 0.05$ bits of positional information, sufficient for each nucleus to determine its location with roughly $1\%$ relative precision and consistent with the measured precision of downstream positional markers \cite{dubuis2013}. Furthermore, this patterning system exhibited signatures suggesting that positional ``gap gene code'' might be optimally organized. This suggests an interesting theoretical program: look  for regulatory network architectures that maximize encoded positional information~\cite{Tkacik_PRE1, Tkacik_PRE2, Tkacik_PRE3, Tkacik_PRE4, Tkacik_PRE5} and compare these \emph{ab initio} predictions to \emph{Drosophila} gap gene data.

Taking a step back from concrete systems that necessarily involve an overwhelming amount of biological detail, there are a number of basic yet still unresolved questions about patterning systems and positional information: How do optimal patterns (i.e., patterns that maximize positional information) look like and what determines their shape? How are efficient patterning strategies different if patterning cues are distributed throughout the domain or are present solely at  domain boundaries? In systems where multiple outputs are simultaneously driven by the same patterning cues, how should these outputs be coupled amongst themselves and across space? Can reliable patterns emerge from very noisy patterning cues, that is, can the readout network actually ``create'' positional information? And finally, what is the interplay between positional information and various aspects of robustness---to noise, to systematic changes in patterning cue levels, or to small variations in system size---that have been extensively discussed in particular biological systems \cite{holloway2006,liu2013,staller2015}?

To address these questions as clearly as possible in a rigorous information-theoretic framework, we follow the methodological approach taken by Wolpert in describing his French Flag model. We start with the simplest toy model of patterning, where smoothly varying patterning cues, e.g., morphogens, drive the expression of ``binary genes'' that only can have two states, {\tt ON} or {\tt OFF}. Clearly, this is not an appropriate assumption for many real patterning systems that rely on intermediate levels of gene expression. Conceptually, however,  this assumption has three major advantages: first, it will provide us with basic theoretical insights that generalize to more complex setups; second, we will be able to easily visualize binary gene expression patterns; and third, we will be able to count the number of distinguishable gene expression states. The latter property is essential to gain an intuitive interpretation of positional information, which is generally measured in an abstract ``currency'' of bits. 

In the following, we start by introducing our minimal 1D model of patterning, which is closely linked to Ising models in statistical physics, where magnetic spins (analogous to our binary {\tt ON}/{\tt OFF} genes) respond to spatially inhomogeneous magnetic field (analogous to our smoothly varying morphogen profile). We briefly review the information-theoretic foundations of positional information relevant to the proposed model. We then systematically explore optimal patterns and how they depend on the shape of the input gradient, noise level, the strength of gene-gene and spatial interactions, etc; as a result, we will be able to give a full account of how these factors affect positional information within our model class. %
\section{Results}

\subsection{A minimal model for a pattern-forming system}

\begin{figure}
\includegraphics[width=8cm]{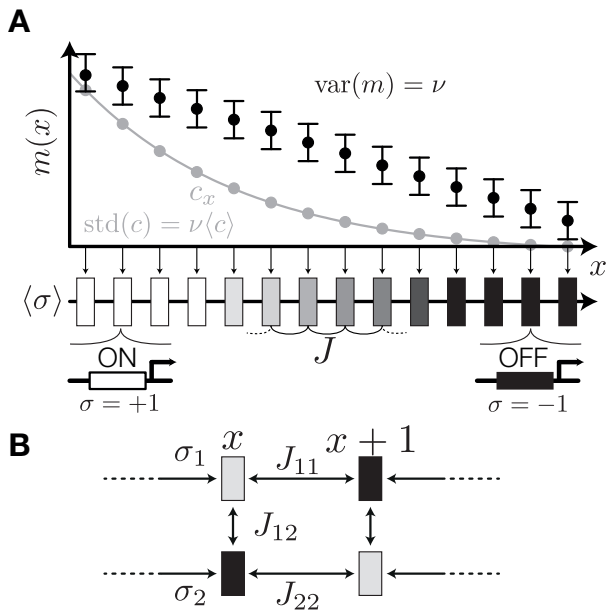}
\caption{
{\bf A schematic diagram of the model system.} {\bf (A)} At each lattice site, $x=1,\dots,N$, we assume a morphogen signal $m(x)$ (black), with extrinsic Gaussian fluctuations of constant variance, ${\rm var}(m) = \nu$ (black error bars). Indicated in gray is the corresponding exponential concentration gradient, $c(x)$, when $m(x)$ is interpreted in the context of  thermodynamic models of gene regulation (see text). The morphogen signal $m(x)$ regulates a binary patterning gene $\sigma(x)$, whose expression state also depends on spatial interaction with strength $J$. Levels of gray denote the mean value $\langle \sigma(x) \rangle$. Hence, $\sigma = +1$ (white) and $\sigma = -1$ (black) mark positions where the gene expression state is deterministic, while levels of gray correspond to noise-induced fluctuations in expression state, with $\sigma=0$ (medium gray) marking positions at which the two expression states are equiprobable. {\bf (B)} For two or more genes per lattice site the model is extended with pairwise local interactions between genes at each lattice site. Spatial interactions are considered only between the same genes at different lattice sites.
}
\label{fig1}
\end{figure}

We look for a model patterning system in which we can systematically explore the effects of gene-gene interactions, spatial interactions, and noise. To keep the task conceptually clean and computationally tractable, we sought for the simplest possible model: we focused on binary ({\tt ON}/{\tt OFF}) patterning genes in the established framework of Ising-like spin models. These paradigmatic systems of statistical physics are \emph{minimal}, i.e., able to generate the relevant phenomenology while having the smallest number of parameters \cite{schneidman2006,tkacik2013}. Our use of such models does not imply that all patterning genes should be viewed as binary (e.g., previous analysis suggests otherwise for \emph{Drosophila} gap genes~\cite{dubuis2013}), or that patterning happens at thermal equilibrium. Nevertheless, conclusions obtained in the simple Ising model framework often do  generalize qualitatively to more complex systems and focus attention on important quantities. At the same time, the Ising framework we introduce below can be seen as a direct extension of Wolpert's original model of binary genes responding to smooth morphogen gradients.

We start with a discrete one-dimensional lattice of $N$ sites, $x=1,...,N$. At every site $x$, the expression pattern is described by a binary variable $\sigma(x)$, with $\sigma(x)=1$ denoting that the patterning gene at location $x$ is {\tt ON}, and $\sigma(x)=-1$ denoting that the gene is {\tt OFF} (see Fig.~\ref{fig1}A). Central to our analysis is the fact that the patterning system is noisy, due to, e.g., intrinsic stochasticity in gene regulation or extrinsic variability in system parameters. To capture the probabilistic nature of the patterning outcomes, we think of the patterning system as generating different spatial patterns $\vec{\sigma} = \{ \sigma(x) \}$ with different probabilities, and in the Ising model framework the probability of each pattern can be written as follows:
\begin{equation}
Q_{\theta}(\vec{\sigma}) = \frac{1}{Z_{\theta}} e^{-H_{\theta}(\vec{\sigma})/\eta}\label{isingprob}\,.
\end{equation}
Here, $Z$ simply ensures that the distribution $Q$ is normalized,  $\eta$ sets the intrinsic noise in the system, and the ``energy,''  $H_{\theta}(\vec{\sigma})$, describes the effect of morphogens and gene-gene interactions on the resulting gene expression pattern. This distribution over all spatial patterns, $Q_{\theta}(\vec{\sigma})$, is parameterized by $\theta$, a set of parameters that we will explicitly identify for our proposed model later. For the ``energy'' $H$, we write:
\begin{equation}
H_{\theta}(\vec \sigma) = -\sum_{x=1}^N h(m(x)) \sigma(x) - J \sum_{x=1}^{N-1} \sigma(x) \sigma(x+1)\label{hamiltonian} \,.
\end{equation}
Here, spatial interaction is modeled by coupling of nearest neighbor sites with a coupling strength $J$. Positive $J$ favor neighboring genes to have equal expression states, whereas negative $J$ favor neighboring genes to have opposing expression states. Biologically, positive $J$ could be realized by diffusion or active transport of gene products between neighboring cells or nuclei; negative $J$, corresponding to repressive spatial interactions, could be mediated by cell-cell signaling networks, e.g., the Delta-Notch pathway \cite{heitzler1991,henrique1995}.

The first term in Eq~(\ref{hamiltonian}) contains the ``bias,'' $h$. The bias favors each individual gene to be either {\tt ON}, whenever $h(x)>0$ at that gene's location, or {\tt OFF}, whenever $h(x)<0$. This term depends explicitly on the coordinate, $x$, and thus models the effect of a morphogen at each location. We can gain biological realism and allow later extensions of the model to multiple patterning genes if we assume that at every location there is a particular value of an abstract morphogen ``signal,'' $m(x)$, which determines the bias in a linear fashion, $h(x) = n (m(x) - E)$. If the signal, $m(x)$, is interpreted as the logarithm of the morphogen concentration, $m(x)=\log(c(x))$, there is an exact mathematical relation between the probability that the patterning gene is {\tt ON}, $P(\sigma(x)=1)$, and the Hill-type thermodynamic model of regulation for the gene $\sigma$. Suppose that the gene $\sigma$ is regulated in a strongly cooperative manner by $\tilde{n}$ binding sites with equal affinities, $\tilde{K}$. Then
\begin{equation}
P(\sigma(x)=1) = \frac{[c(x)]^{\tilde{n}}}{[c(x)]^{\tilde{n}} + K^{\tilde{n}}},
\end{equation}
and it is easy to show that the parameters $n$ and $E$ relating the bias $h(x)$ to the morphogen signal $m(x)$ in our model correspond, up to a multiplicative factor, to $\tilde{n}$ and $\log(\tilde{K})$ in the thermodynamic model of regulation.
Furthermore, observed morphogen concentration gradients that commonly have an exponential profile, $c(x)=c_0\exp(-\lambda x)$, map to linear morphogen signals, $m(x)=-\lambda x+\log(c_0)$, in our framework. Linear $m(x)$ is thus our baseline profile, although we will subsequently also define and explore more localized morphogen signals. 

In our model, $\eta$ controls intrinsic fluctuations in the system. For $\eta\rightarrow 0$, the gene $\sigma$ responds deterministically to the morphogen and the expression state at neighboring locations; for example, if there were no spatial interactions ($J=0$), the gene would be {\tt ON} whenever the local morphogen signal exceeds the threshold, $m(x)>E$, and {\tt OFF} otherwise, following Wolpert's original idea. In contrast, for $\eta\rightarrow \infty$ the genes respond completely randomly, with equal probability of being {\tt ON} or {\tt OFF}, irrespective of the relevant signals. In addition to this \emph{intrinsic noise}, we also consider \emph{extrinsic noise}~\cite{swain2002}. To that end, we assume that there can be fluctuations in the morphogen signal, $m$, that are additive and Gaussian with constant variance $\nu$, i.e., ${\rm var}(m)=\nu$. This maps to fluctuations in morphogen concentration that are proportional to the mean concentration, ${\rm std}(c) = \nu \langle c \rangle$, a dependence that is biophysically plausible and has previously been discussed in the literature \cite{swain2002,paulsson2004}.

An extension of the model from a single patterning gene to multiple genes is straightforward (see Fig.~\ref{fig1}B). Let there be $K$ distinct patterning genes, such that at each location $\boldsymbol{\sigma}(x) \equiv \{\sigma_\alpha(x)\}$, for $\alpha=1,\dots,K$. To write down the function $H_\theta(\vec{\boldsymbol{\sigma}})$ and compute the probability of every pattern, we simply reproduce the ``bias'' and spatial interaction terms for each one of the $K$ genes and sum them up in the energy function. Next, we add a qualitatively new term that couples the $K$ genes amongst themselves at every location $x$, which models activating ($J_{\alpha\gamma}>0$) or repressive ($J_{\alpha\gamma}<0$) interactions between the genes $\alpha$ and $\gamma$ (where $\alpha,\gamma\in \{1,\dots,K\}$ and $\alpha \neq \gamma$). The complete energy function  reads:
\begin{eqnarray}
H_{\theta}(\vec{\boldsymbol{\sigma}}) &=& -\sum_{\alpha=1}^K \sum_{x=1}^N h_\alpha(m(x)) \sigma_\alpha(x) - \nonumber \\
&-& \sum_{\alpha=1}^K J_{\alpha\alpha} \sum_{x=1}^{N-1}  \sigma_{\alpha}(x) \sigma_{\alpha}(x+1) - \nonumber \\
&-&   \sum_{\substack{\alpha,\gamma=1\\\alpha\neq\gamma}}^K J_{\alpha \gamma} \sum_x \sigma_{\alpha}(x) \sigma_{\gamma}(x) \label{hamiltonianfull} \,.
\end{eqnarray}
This model of $K$  genes responding to a morphogen signal is thus fully specified by $3K + K(K+1)/2$ interaction parameters $\theta=\{ n_\alpha, E_\alpha, J_{\alpha\alpha}, J_{\alpha\gamma}\}$, the intrinsic noise $\eta$, the extrinsic noise $\nu$, and the shape of the morphogen gradient, $m(x)$. These parameters, together with Eqs~(\ref{isingprob},\ref{hamiltonianfull}), fully specify the resulting distribution over gene expression patterns, $Q_\theta(\vec{\boldsymbol{\sigma}})$. Next, we formally define positional information and discuss its computation and behavior within our model.

\subsection{Positional information for genes with two possible expression states}

We use a previously introduced information-theoretic definition of positional information~\cite{dubuis2013,tkacik2015}. This definition is based on the observation that a cell can only ``know'' as much about its location $x$ as it can infer from local and generally noisy expression levels, $\boldsymbol{\sigma}(x)$, of patterning genes. Again, $\boldsymbol{\sigma}(x)$ denotes the vector of all considered gene expression states at location $x$. Let the distribution of $\boldsymbol{\sigma}$ at location $x$ be $P(\boldsymbol{\sigma}|x)$. Positional information is then defined as the mutual information between the location $x$ and the local expression level $\boldsymbol{\sigma}$:
\begin{equation}
I(\boldsymbol{\sigma};x)=S[P_\sigma(\boldsymbol{\sigma})] - \langle S[P(\boldsymbol{\sigma} | x)]\rangle_x \,. \label{info1}
\end{equation}
The first term in Eq~(\ref{info1}) is the entropy of the distribution of expression states across the lattice, with $P_{\sigma}(\boldsymbol{\sigma}) = \frac{1}{N} \sum_{x=1}^N P(\boldsymbol{\sigma} | x)$, which favors diverse use of expression states across the spatial pattern. The second term is a penalty term that quantifies the average variability in gene expression which is uncorrelated with position. Because it can induce confusion in the expression-position mapping, this ``noise entropy'' can only reduce the information and is zero in a noiseless system.

Considering a single patterning gene without noise, gene expression achieves maximum positional information by partitioning all $N$ locations into two equally sized sets, one where the gene is {\tt OFF} and one where the gene is {\tt ON}. Without noise, the second term in Eq~(\ref{info1}) vanishes and the positional information is given by the entropy of expression states. For a balanced assignment of expression states to positions we have $P_{\sigma}(\sigma=+1)=P_{\sigma}(\sigma=-1)=\frac{1}{2}$ and thus $S[P_{\sigma}(\sigma)]=1$. Such a pattern is said to convey ``one bit'' of positional information, an amount sufficient to reliably separate the anterior from the posterior, or odd from even rows of cells; generally, one bit corresponds to the amount gained by an unambiguous answer to an optimally posed yes/no question. In the case of multiple patterning genes, a combination of gene expression states, e.g., $\{\mathtt{ON, OFF, ON, ON}\}$ or $\{+1, -1, +1, +1\}$, can  be seen as a ``codeword'' for some particular position. The capacity of this code, independently of how the gene expression patterns are set up or read out, is again given by the first term in Eq~(\ref{info1}). For four binary genes this cannot exceed 4 bits; as in the case of a single gene, the bound is achieved when each of the $2^4=16$ distinct gene expression combinations is used equally often across all locations $x$ in the tissue.

Mutual information is symmetric in its arguments, so that we can rewrite Eq~(\ref{info1}) as
\begin{equation}
I(\boldsymbol{\sigma};x) = S[P_x(x)] - \langle S[P(x| \boldsymbol{\sigma})]\rangle_{\boldsymbol{\sigma}},  \label{infodecoding}
\end{equation}
where $S[P_x(x)] = \log_2 N$ for cells that are uniformly distributed over coordinate $x$, as we assume in our model. This way of writing positional information emphasizes that $\log_2(N)$ bits is the upper bound on the positional information in a patterning system, which would correspond to an unambiguous identification of every location $x$ based on the expression pattern $\boldsymbol{\sigma}$. Positional information is decreased from this bound by the second term in Eq~(\ref{infodecoding}) which measures the uncertainty in position that remains even when one knows gene expression levels. Indeed, one can be more explicit about the error in trying to estimate the position, $x$, given the expression levels $\boldsymbol{\sigma}$. It can be shown (see S1 Appendix 1) that the expected error of any estimator $\hat{x}$ for position is bounded by positional information:
\begin{equation}
\mathbb{E}(x-\hat{x})^2 \geq \frac{1}{12}\left(N^2 2^{-2I(\boldsymbol{\sigma};x)}-1\right). \label{errorbound}
\end{equation}
The bound on the error can be made vanishingly small if positional information approaches its upper bound, $I\rightarrow \log_2N$ bits. This relationship is analogous to the relationship between positional error and positional information for continuous systems~\cite{dubuis2013,tkacik2015}. Importantly, when cells need to make decisions appropriate to their position within an organism, they are also subject to the  estimation limits of Eq~(\ref{errorbound}), irrespective of how complex the molecular readout mechanism for $\boldsymbol{\sigma}$ is.

How is the pattern-forming process $Q_\theta(\vec{\boldsymbol{\sigma}})$  related to $P(\boldsymbol{\sigma}|x)$, which determines positional information? While the pattern-forming process $Q$ yields a global (joint) distribution over all possible patterns, the positional information is local and thus is only concerned with the expression state at a given location $x$; thus, $P(\boldsymbol{\sigma}|x)$ is obtained by summing (marginalizing) the joint distribution over gene expression states everywhere but at $x$:
\begin{equation}
P(\boldsymbol{\sigma}|x) = \sum_{\substack{\boldsymbol{\sigma}(x') =\pm 1\\ x'\neq x}}Q_\theta(\vec{\boldsymbol{\sigma}}). \label{marginalization}
\end{equation}
Therefore, we can use Eqs~(\ref{isingprob},\ref{info1},\ref{marginalization}) to compute positional information, $I(\boldsymbol{\sigma};x)$, which we will refer to as ``PI'' in the text, for any patterning system with parameters $\theta$. Using standard approaches from statistical physics (transfer matrices) these mathematical manipulations are all doable exactly when the extrinsic noise, $\nu$, is zero. When $\nu\neq 0$, analytical techniques coupled with tractable Monte Carlo sampling can be used to compute PI as a function of parameters; see S2 Appendix 2 for details.

Our framework clearly separates the pattern-forming process whose outcome is described by $Q_\theta(\vec{\boldsymbol{\sigma}})$ and that depends on mechanistic parameters $\theta$, from the resulting pattern, which carries a certain amount of positional information $I(\boldsymbol{\sigma};x)$. Such a distinction is important and clarifies a number of conceptual issues with positional information. Consider, for example, the case of a single, noiseless gene responding to a smoothly varying morphogen gradient. As explained above, the maximal positional information achievable in this system is 1 bit. On the other hand, one could argue that by changing the threshold at which the readout gene is activated, the system can position the pattern boundary in any of $N+1$ possible locations (including the system boundaries), generating $N+1$ possible patterns: shouldn't the information then be $I=\log_2(N+1) \geq 1$ bits? The apparent contradiction is that $\log_2(N+1)$ bits is \emph{not} positional information; rather, it is a measure of how many different joint patterns, $\vec{\boldsymbol{\sigma}}$, can be generated by changing the parameters $\theta$ of the pattern forming system (in this case the readout threshold), i.e., the mutual information $I(\vec{\boldsymbol{\sigma}};\theta)$, which is  an intrinsic property of the pattern-forming process, $Q$. This quantity certainly affects positional information, $I(\boldsymbol{\sigma};x)$, yet it is not identical to it. In particular, positional information can be different for each separate value of parameters $\theta$. As we will see in our toy model, one can find patterning systems that have a high positional information while simultaneously having either a high or low value for $I(\vec{\boldsymbol{\sigma}};\theta)$, and vice versa. The two information-theoretic quantities therefore describe independent aspects of the system, and should not be confused with each other.

\begin{figure*}
\includegraphics[width=12cm]{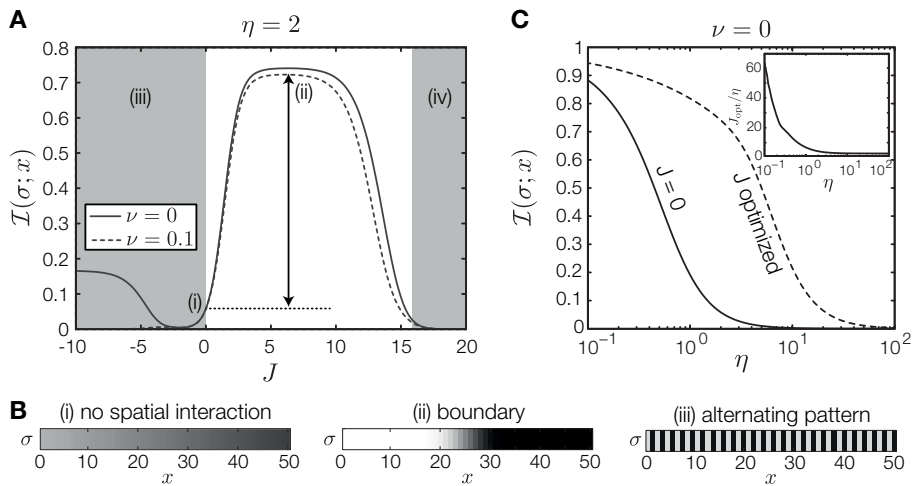}
\caption{
{\bf Effect of spatial interactions on one patterning gene.} {\bf (A)} PI as a function of spatial interaction strength $J$ with fixed intrinsic noise $\eta=2$ and two levels of extrinsic noise (legend). The arrow indicates the excess PI available to an optimally spatially coupled system, relative to an uncoupled ($J=0$) system. At nonzero extrinsic noise ($\nu=0.1$) PI is strongly suppressed at $J<0$. {\bf (B)} The average spatial pattern for three regions denoted in (A): (i) no spatial interaction; (ii) positive $J$ stabilizes a pattern with a boundary against noise; (iii) negative $J$ results in an alternating pattern. In region (iv) indicated in (A) the strength of spatial interactions forces the pattern in a uniform all {\tt ON} or all {\tt OFF} state that carries 0 bits of positional information. {\bf (C)} Comparison between PI with (dotted line) and without (solid line) spatial interaction as a function of intrinsic noise, $\eta$. For the spatially coupled system an optimal $J_{\rm opt}$ has been found separately for each value of intrinsic noise $\eta$. The corresponding values of $J_{\rm opt}$, scaled by the respective noise levels $\eta$, are plotted in the inset.
}
\label{fig2}
\end{figure*}

\subsection{Optimal patterning with a single gene depends on noise level and morphogen profile}

\begin{figure*}
\includegraphics[width=12cm]{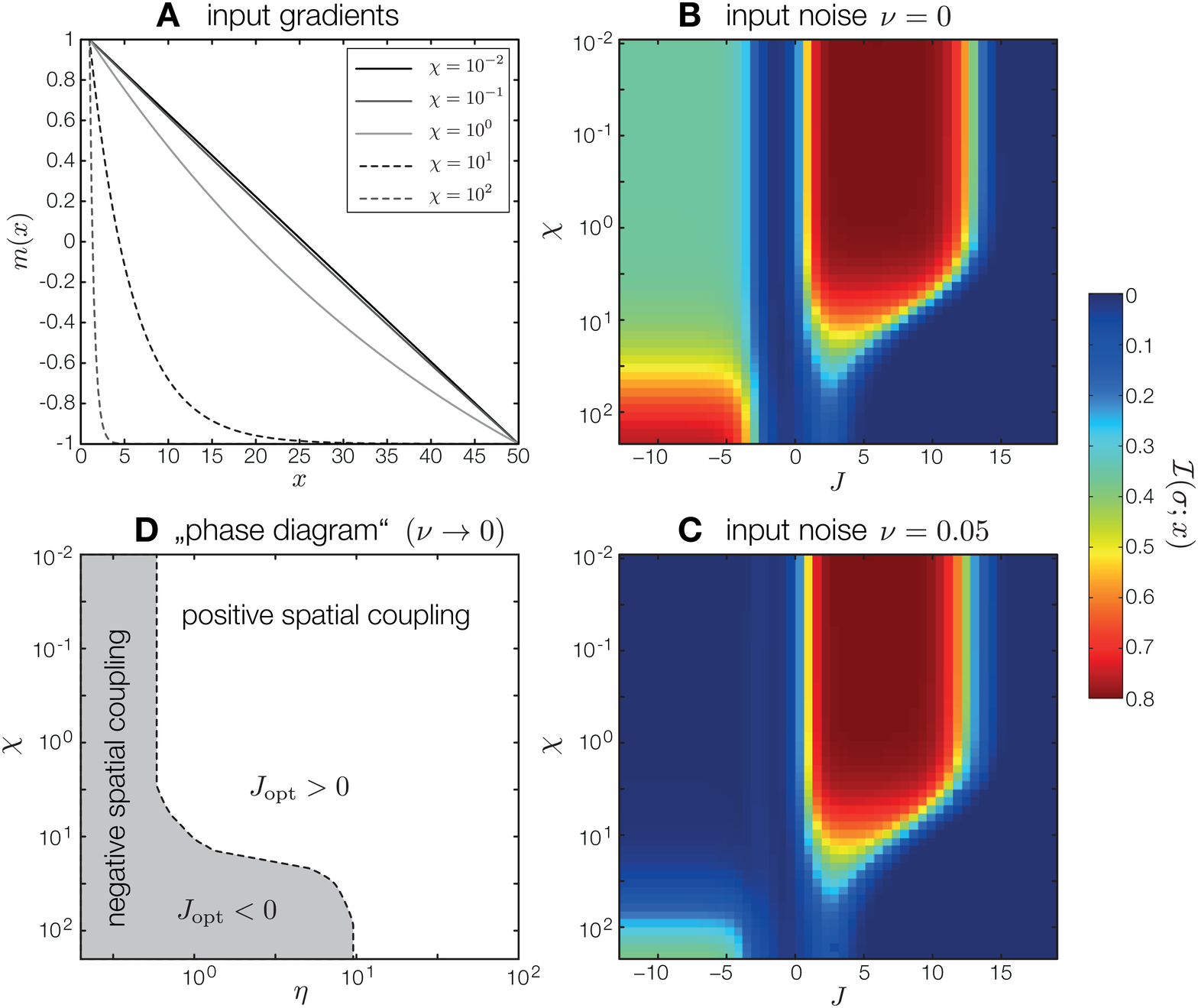}
\caption{
{\bf Effect of gradient shape on PI carried by one patterning gene.} {\bf (A)} Morphogen signal profiles, $m_{\chi}(x)$, for different shape parameters, $\chi$, interpolate between a linear profile and a profile strongly concentrated at the anterior boundary. {\bf (B)} PI carried by one patterning gene as a function of spatial interaction strength $J$ and shape parameter $\chi$. The intrinsic noise is set to $\eta = 1.25$, with zero extrinsic noise. {\bf (C)} Same as (B), with extrinsic noise added. While PI in the regime of positive spatial interactions is almost unchanged, PI for negative spatial interactions is greatly diminished. {\bf (D)} Phase diagram (at $\nu=0$) depicting for which values of intrinsic noise and gradient shape negative or positive coupling (and thus the resulting boundary or alternating pattern) is optimal.
}
\label{fig3}
\end{figure*}

We start by studying a  system with one patterning gene, $\sigma$, on a lattice of $N=50$ sites. We choose a linearly decaying morphogen signal, $m(x)$, which favors the {\tt ON} state of the patterning gene in the anterior and {\tt OFF} state in the posterior. In absence of any noise and spatial interactions, we have a clear expectation for the pattern that yields the maximally achievable PI of 1 bit: this will be a symmetric partition of the lattice into equally sized anterior ({\tt ON}) and posterior ({\tt OFF}) halves, with a single sharp boundary at $x=N/2$. In terms of the parameters of our energy function, Eq~(\ref{hamiltonian}), this corresponds to choosing $E=J=0$, $n=1$ and $\eta,\nu \rightarrow 0$. 

How does the spatial interaction $J$ affect the ability of a single gene $\sigma$ to encode positional information when noise is not zero? Figure~\ref{fig2} shows $I(\sigma;x)$ as a function of the coupling strength $J$ at fixed levels of intrinsic ($\eta$) and extrinsic ($\nu$) noise. In the absence of spatial interactions ($J=0$) the response of the gene $\sigma$ is uncoordinated across positions and gets largely destroyed by random fluctuations (Fig.~\ref{fig2}B(i)). Consequently, PI is low, in this case below 0.1 bits.


If spatial interaction is increased to positive values of $J$, PI increases steeply to a maximum. The emergent pattern is qualitatively consistent with the expected optimal pattern: it contains a single boundary that divides the lattice into anterior and posterior halves (Fig.~\ref{fig2}B(ii)). The effects of noise on the patterning gene are restricted to the area around the boundary. Therefore, the maximally achievable PI is determined by the accuracy with which the boundary is positioned and depends on the noise level. For this optimal boundary pattern, additional extrinsic noise, $\nu$, diminishes PI only mildly. 

The large increase in PI can be ascribed to the well-studied effect of spatial averaging \cite{erdmann2009, Tkacik_PRE4}. With increasing coupling strength $J$, different lattice sites do not respond independently anymore but are spatially correlated. Consequently, fluctuations at individual lattice sites are overcome by a concerted response to the input field integrated over a range, which greatly sharpens the resulting pattern. As $J$ is increased further, PI decreases again and eventually drops to zero because the lattice is forced into a spatially uniform (all {\tt ON} or all {\tt OFF}) configuration. Figure~\ref{fig2}C summarizes the benefit of  positive spatial interactions. The solid curve is the PI without spatial interactions plotted as a function of intrinsic noise $\eta$, whereas for the dashed curve $J$ has been optimized separately for each value of $\eta$. The values of the optimized $J$ as a function of $\eta$ are shown in the inset. The spatially coupled system is therefore capable of retaining PI above 0.5 bits for more than an order of magnitude higher internal noise relative to the system without spatial interaction. In sum, positive spatial coupling $J$  has the ability to stabilize the near optimal pattern against effects of intrinsic and extrinsic noise.

A qualitatively different pattern is formed if $J$ is decreased to negative values. Initially, for small negative $J$, PI decreases almost to zero as negative spatial interaction disturbs the readout of the morphogen signal. Beyond the minimum, PI increases again and the system generates a pattern in which neighboring lattice sites alternate between {\tt ON} and {\tt OFF} states (Fig.~\ref{fig2}B(iii)). This is an alternative strategy for encoding PI: the system now distinguishes between even and odd lattice positions instead of an anterior and a posterior segment. While in principle both patterns can encode a full bit of PI, we will see that the alternating pattern is less robust against noise. 


For the alternating pattern to form correctly it is necessary that the {\tt ON}/{\tt OFF}  sequence is faithfully propagated through the bulk and that the morphogen signal can reliably break the symmetry between the two possible alternating patterns (if both patterns are equally likely,  PI goes to zero). The condition for robust propagation depends on  intrinsic noise. Breaking the symmetry is, on the other hand, highly susceptible to extrinsic noise, i.e., fluctuations in the morphogen signal. Consider, for example, the case of the linear gradient, which exerts the strongest bias at the anterior- and posterior-most sites. The ability to reliably break the symmetry depends on the anterior site consistently experiencing a stronger bias towards {\tt ON} than the second site (which in an alternating pattern should be {\tt OFF}), i.e., the mean difference in the morphogen signal between the first two sites needs to be larger than the typical strength of  extrinsic fluctuations in the signal, $ \langle m(x=1)-m(x=2) \rangle > \nu$, otherwise  PI of the alternating pattern will be severely impaired. A similar argument can be made for the influence of  intrinsic noise. Even if spatial interaction is strong enough to allow only strictly alternating patterns, it is still possible that the entire pattern flips to its inverse due to intrinsic fluctuations. Again, the ability of the system to select between the two possible patterns depends on the difference of the mean values at neighboring lattice site compared to the intrinsic noise level.


How does the optimal strategy for a single patterning gene depend on the shape of the morphogen signal? To investigate this, we consider a set of exponential shapes for the morphogen signal $m(x)$, parametrized by a a decay parameter, $\chi$. For $\chi\ll 1$, we recover the linear morphogen signal discussed above, while for larger $\chi$ the morphogen signal is increasingly concentrated at the anterior, as shown in Fig.~\ref{fig3}A. Figure~\ref{fig3}B shows PI carried by a system with one gene as a function of $\chi$ and coupling strength $J$. Because localized morphogen signals can reliably break the symmetry of an alternating pattern, high PI can be achieved with a combination of negative $J$ and large $\chi$. In contrast, patterns that form a boundary (with $J>0$) are efficient if the morphogen signal extends throughout the system. Consistent with our first observation, a systematic survey in Fig.~\ref{fig3}C shows that small additions of extrinsic noise severely lower  PI carried by the alternating pattern while leaving the boundary pattern almost unaffected. 

These observations can be summarized in a phase diagram, shown in Fig.~\ref{fig3}D. The diagram, constructed for zero extrinsic noise ($\nu=0$), divides the plane spanned by intrinsic noise $\eta$ and morphogen shape parameter $\chi$ into a region where negative spatial interaction is optimal and a region where positive spatial interaction is optimal. For  very low intrinsic noise, the alternating pattern generally outperforms the boundary pattern. As  intrinsic noise increases, the optimal patterning strategy depends on the form of the gradient: for a spatially extended gradient the boundary pattern is optimal, whereas for a gradient concentrated at the boundary the alternating pattern is optimal. For sufficiently high intrinsic noise, the boundary pattern always outperforms the alternating pattern, even for gradients concentrated at the boundary. Adding extrinsic noise generally shifts the boundary in the phase diagram to favor positive spatial interactions.

\begin{figure*}
\includegraphics[width=15.5cm]{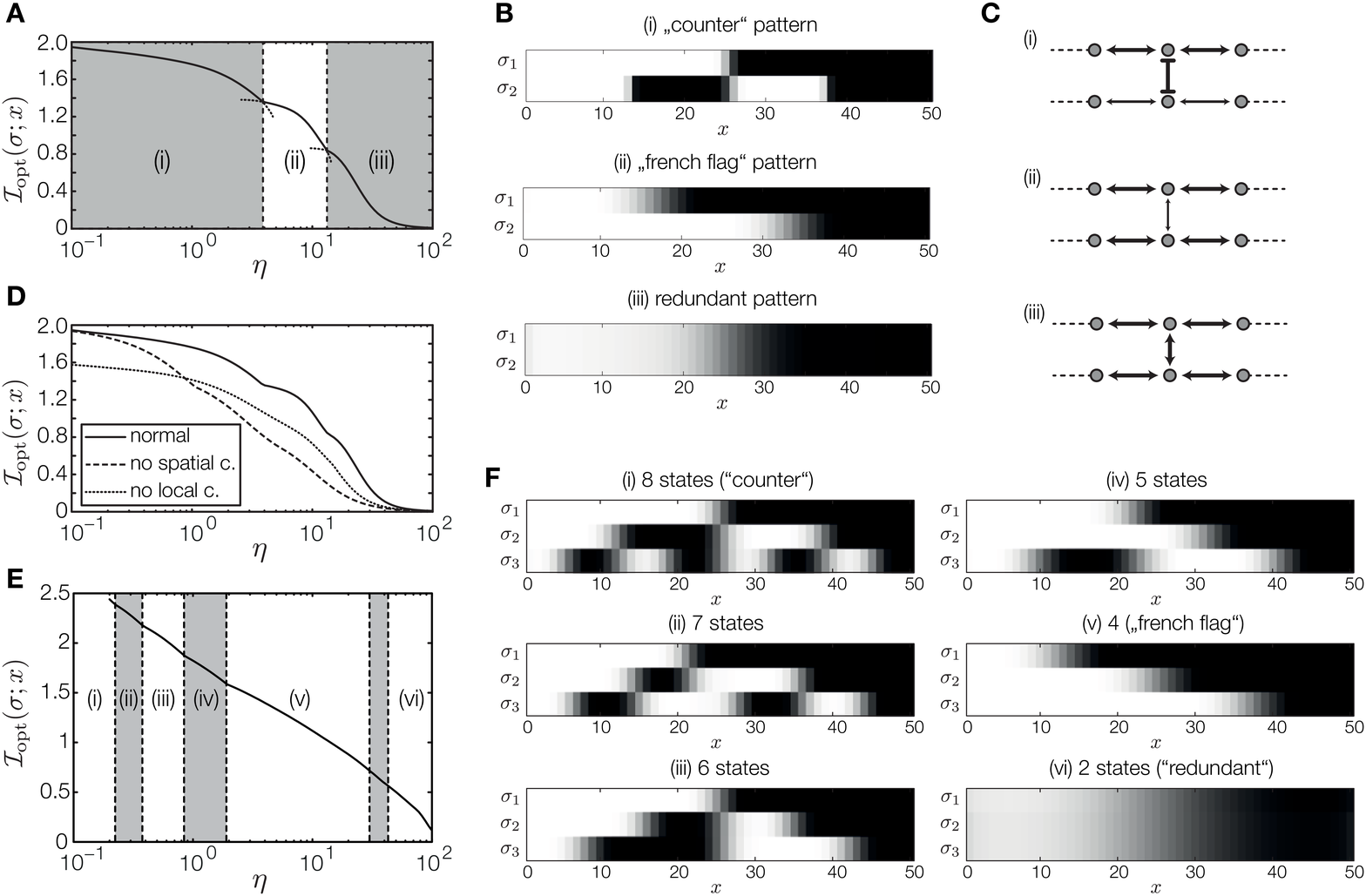}
\caption{
{\bf Positional information carried by two and three patterning genes with a linear morphogen signal.} {\bf (A)} PI as a function of intrinsic noise level. For each noise level $\eta$, all parameters of the network have been optimized. The depicted regions (i)-(iii) indicate different numbers of states encoded by the resulting patterns. At each boundary (non-optimal) continuations of the coding strategy in the neighboring region are depicted as dotted curves. {\bf (B)} Characteristic patterns of the different regions in (A). {\bf (C)} Schematics of the network parameters for the different coding strategies. Pointed arrows denote positive interaction, blunted arrows denote negative interaction. The thickness of the arrows indicates their strength. {\bf (D)} Comparison of PI carried by systems without spatial interactions and systems without local gene-gene interactions as a function of intrinsic noise. For each noise level the parameters of each system have been optimized separately. {\bf (E)} PI carried by three patterning genes as a function of intrinsic noise. {\bf (F)} Characteristic three-gene patterns for the different regions in (E).
}
\label{fig5}
\end{figure*}

\subsection{Optimal patterning with multiple interacting genes can establish a stable combinatorial code for position}

Which patterns optimally encode positional information in systems with multiple patterning genes? 
The French Flag model proposes a cascaded activation of the genes in response to the morphogen signal. For two binary genes this would lead to a pattern with three separate states (the ``Tricolore'' of the French Flag) and a maximal PI of $I(\boldsymbol{\sigma};x)=\log_2(3) \approx 1.59\,\,{\rm bits}$. In contrast, we know from the arguments made above that an optimal pattern with two binary genes should have PI of 2 bits, at least with vanishing intrinsic and extrinsic noise. In such a pattern all possible expression states would be realized and  evenly distributed throughout the lattice. For two binary genes, $(\sigma_1,\sigma_2)$, there are four different possible states: ({\tt ON},{\tt ON}), ({\tt ON},{\tt OFF}), ({\tt OFF},{\tt ON}) and ({\tt OFF},{\tt OFF}); note that one of the mixed states is missing in the French Flag, which is why it has lower PI. We will refer to  PI-maximizing binary patterns for $K$ genes, where all $2^K$ states occur with equal probability in the pattern, as ``Counter'' patterns. A Counter pattern is an example of a \emph{combinatorial code}, where position can only be decoded properly when the readout mechanism has simultaneous and complete access to the local expression states of all $K$ genes.

We can ask two fundamental questions about Counter patterns. First, can such patterns be generated in a model where genes interact locally in a pairwise fashion and are spatially coupled, as assumed by Eq~(\ref{hamiltonianfull})? Second, Counter patterns are clearly optimal when noise is vanishing; are they optimal also when noise is present? If not, what are the optimal patterns in that case?

To investigate these questions we optimize the parameters of our model for two and three genes to find patterns that maximize PI for different levels of noise. Specifically, we vary all interactions in the system, both spatial ($J_{\alpha\alpha}$) and regulatory ($J_{\alpha\gamma}$), as well as the parameters that prescribe how each gene couples to the morphogen signal ($\{n_\alpha,E_\alpha\}$). For the case of two (three) patterning genes, this amounts to a total of 9 (15) parameters; we use stochastic optimization to carry out PI maximization (see S2 Appendix 2 for details). As with the single gene case, we assume a linear morphogen signal, $m(x)$. The only remaining dependence of our results is thus on the strength of the intrinsic ($\eta$) and extrinsic ($\nu$) noise.

When the noise level is low enough, we find that one of the genes always takes on strongly negative spatial interaction, $J_{\alpha\alpha}<0$, to generate an alternating pattern. This gene does not interact with the others, and contributes one bit (in the low noise limit) to the total PI. There can only be one such gene in the pattern, as any subsequent alternating gene (with the same or opposite polarity) would be redundant with the first one and thus provide no further increase in PI. As this strategy to increase PI by one bit is trivially available to any patterning system at low noise and does not occur at high noise, we restricted our subsequent search only to cases where spatial interactions are restricted to be positive, $J_{\alpha\alpha}>0$, mimicking spatial averaging induced by diffusion or active transport of patterning gene products.


Figure~\ref{fig5}A shows the maximal PI carried by two binary genes as a function of intrinsic noise, $\eta$. Characteristic examples of corresponding output patterns are shown in Fig.~\ref{fig5}B.  As expected, the optimal pattern at low noise is a Counter, realizing each of the four possible expression states in an equally sized spatial fraction. A network schematic illustrating the optimized parameters is depicted in Fig.~\ref{fig5}C. As noise increases, the values of optimal parameters undergo an abrupt change and the optimal pattern changes from a Counter to a French Flag, encoding only three states (region (ii)). At the boundary between the two regions there is a visible kink in the PI curve. Continuations of the two coding strategies into the respective other noise regime are depicted as dotted curves in Fig.~\ref{fig5}A. If the noise level is increased even further, the optimal network changes its coding strategy again and generates a pattern in which both genes redundantly form a boundary in the center (region (iii)).

What are the respective influences of spatial and local gene-gene interactions on the formation of patterns and encoding of positional information? To study this question, Fig.~\ref{fig5}D compares the curve of Fig.~\ref{fig5}A with curves for PI carried by systems which are optimized without spatial interactions (dashed curve) and without local interactions (dotted curve). Without local interaction between genes the maximally achievable PI is about 1.59 bits, corresponding to the three-state French Flag pattern. This observation can be generalized: if a monotonous input gradient acts on the target genes \emph{independently} via a monotonous response function (e.g., a sigmoid, as here), then each gene can form only a single transition or boundary. In that case, the French Flag pattern indeed encodes the maximally achievable PI. If, however, the genes interact with each other, their response functions receive multiple inputs and complex patterns are possible. For instance, the system generating the Counter pattern (Fig.~\ref{fig5}C(i)) has strong mutual repression between the two genes, which switches the lower gene to the inverse of the upper gene where the strength of the morphogen signal is low. Systems without spatial interactions, in contrast, can and do carry as much PI as the fully interacting system when the noise is vanishing. As the noise level increases, however, the fully interacting system always outperforms the system without spatial coupling, demonstrating the important role of spatial noise averaging. At high noise, the optimal boundary pattern is identical for both genes, providing two redundant read-outs of the morphogen signal. In this case, the local gene-gene interaction is positive and strong, providing further noise averaging (across the two readout genes), beyond that due to spatial interactions.

\begin{figure*}
\includegraphics[width=13cm]{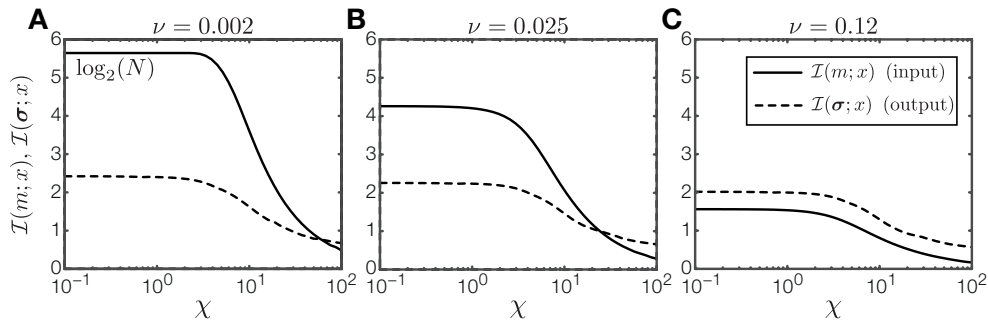}
\caption{
{\bf Comparison of PI in input gradient and output pattern.} PI in $m$ (solid) and $\boldsymbol{\sigma}$ (dashed) are shown for three different input noise levels ((A)-(C)) as a function of $\chi$, parametrizing the gradient shape (cf. Fig.~\ref{fig2}A). The computation of $\mathcal{I}(m;x)$ is described in S3 Appendix 3.
}
\label{fig6}
\end{figure*}

Maximization of PI in a system with three genes corroborates our results for two genes. Again, as intrinsic noise increases, the optimal strategy switches abruptly at particular noise values, marking a transition to a code that specifies one less distinct expression state.  Between transitions, the number of distinct states that the network can generate is held constant, but information nevertheless decreases smoothly as increasing noise leads to more ambiguity in the mapping between position and gene expression state (Fig.~\ref{fig5}E). Three binary genes can  generate a maximum  of eight distinct expression states, which is achieved by a three gene Counter pattern (Fig~\ref{fig5}F(i)). Between the optimal Counter pattern and the fully ``redundant'' pattern encoding two states, we find a variety of other strategies, including  French Flag,  that specify an intermediate number of states (Figs.~\ref{fig5}F(ii)-(iv)). The Ising-model-based framework for binary interacting genes is clearly sufficiently rich to generate this variety, including the theoretically optimal Counter. It is possible that even richer models, e.g., models allowing three-way interactions between genes in addition to pairwise interactions, would lead to higher PI values, possibly by being more robust to noise, or by allowing the Counter pattern to be easily generated in systems with $K>3$ genes.

These results are not changed significantly by the addition of extrinsic noise. Generally, increasing extrinsic noise decreases the maximum achievable PI and can, analogously to intrinsic noise, lead to a change of the optimal patterning strategy. The effects of extrinsic noise can be effectively attenuated by spatial interaction of the patterning genes. This raises the question of how much positional information loss due to fluctuations in the input morphogen can be avoided by a spatially interacting network of patterning genes.

\subsection{Positional information of spatially interacting patterning genes can exceed that of the morphogen signal}
Can  patterning genes encode more positional information than the morphogen signal itself? In other words, can a network generate PI starting with a noisy  signal? Intuitively, both Turing patterns and cellular automata would suggest that the answer to this question is affirmative. In a Turing model at steady state, spatial locations are assignable to (at least) two expression states, so that there is positional information where initially there was none.  Similarly, consider the simplest cellular automaton that proceeds along one discrete spatial dimension with a simple rule: \emph{``Read the value in the current position, increment the value by one, move one cell to the right, write the value.''} Such a cellular automaton would generate a separate cell fate (i.e., a unique numerical value)  in each position, providing maximal PI. In both cases, PI before patterning appears to be zero, and after patterning has some nonzero value. 

More careful thought reveals, however, that the PI was established by transforming information that must have been present already at the beginning of the patterning process.  Turing patterning is a deterministic mechanism defined by a set of partial differential equations, whose steady state solution therefore depends on the initial condition and the shape of the boundary. While the mechanism will generically produce domains with a typical lengthscale separated by sharp borders, the positions of the borders will shift with the initial and boundary conditions. Specifying initial and boundary conditions, however, requires information: more bits for more precise specification. Although the formal link between this information and the resulting PI depends on the system, it is clear that an ensemble of Turing systems with arbitrary initial / boundary conditions will not yield an ensemble of patterns with appreciable PI. This is even more obvious in the cellular automaton example. To apply the proposed rule and generate the pattern, one needs to specify the initial condition: the numerical value in the cell at position one. Suppose that there is $N$ possible lattice positions. Specifying the initial value, $\theta_0$, for position one then amounts to providing $I_0=\log_2 N$ bits of information for the automaton to start working. The PI of this initial pattern (with the first cell specified, and all others unspecified) is low. After the automaton finishes, all $N$ positions are uniquely specified, yielding PI of $I(\sigma;x)=\log_2 N$ bits. The automaton has therefore taken the initial information $I_0$ and ``spread it over space'' to generate PI of equal amount. If, however, the initial value is specified poorly (i.e., probabilistically, with close to uniform distribution), the resulting ensemble of patterns will have very small PI.

These restrictions trace back to the Data Processing Inequality (DPI) \cite{cover2012}, a central result in information theory. If initial/boundary conditions for a deterministic patterning process are specified with finite precision (e.g., they are noisy across repetitions of the same pattern generation), then the resulting patterns can again be seen as draws from the distribution $Q_\theta(\vec{\boldsymbol{\sigma}})$, where $\theta$ are now interpreted as the true boundary / initial conditions, which, however, enter the patterning dynamics with some noise. This is much like the extrinsic noise that corrupts the morphogen signal in our Ising-like model. In the case of the simple cellular automaton, it is easy to convince oneself that the final PI must be equal to the initial information, $I_0$. The corresponding patterning can be seen as a Markov chain: $\theta_0\rightarrow\tilde{\theta}_0\rightarrow\vec{\sigma}$, where $\theta_0$ is the ``true'' initial condition, $\tilde{\theta}_0$ is its corrupted version, which corresponds to the initial condition not being specified with perfect precision, and $\vec{\sigma}$ is, as before, the resulting pattern. In this case, DPI states that $I(\theta_0; \vec{{\sigma}}) \leq I(\theta_0;\tilde{\theta}_0)$---the precision by which the initial state is specified limits the reproducibility of the resulting pattern. Since in this example $I(\theta_0; \vec{{\sigma}}) = I({\sigma};x)$, PI is limited by the specification of the initial state. The same type of argument applies generally, although the bounds for more complex systems may be difficult to derive. 

What limits does the Data Processing Inequality imply for our model system? First, unlike the cellular automaton and the Turing mechanism examples above, our patterning takes place at equilibrium, so that the dependence on  initial conditions is lost. When spatial interactions in our model are set to zero, the expression states at individual locations of the lattice become independent of each other. The patterning can then be seen as a Markov chain, $x\rightarrow m\rightarrow\boldsymbol{\sigma}$, and DPI requires that $I(\boldsymbol{\sigma};x)\leq I(m;x)$, i.e., that PI of the patterning genes must be smaller or equal to PI of the morphogen signal. In this case, the network, no matter how complicated, cannot provide more PI than the morphogen signal already has. 

How does this picture change when we allow spatial interactions? Figure~\ref{fig6} compares PI carried directly by the morphogen signal with PI carried by an optimized network of three patterning genes responding to that morphogen signal. For steep gradients or high extrinsic noise, PI of the patterning genes can indeed exceed PI in the morphogen signal itself, which appears to be in violation of  DPI. 

To explain this observation, we turn to the theoretical question of whether a patterning network downstream of the morphogen signal can contribute to encoding of PI beyond the information already present in the  signal. More specifically, given a morphogen signal $m$ and several patterning genes $\boldsymbol{\sigma}$ responding to it, is it possible that $I(\boldsymbol{\sigma},m;x) > I(m;x)$? Here, $I(\boldsymbol{\sigma},m;x)$ is PI  jointly carried by the simultaneous state of both $\boldsymbol{\sigma}$ and $m$ about position $x$. Joint information can be split up as follows: 
\begin{equation}
I(\boldsymbol{\sigma},m;x) = I(m;x) + I(\boldsymbol{\sigma};x|m)\,,
\end{equation}
where the last term is the conditional mutual information
\begin{equation}
I(\boldsymbol{\sigma};x|m) = \left\langle  \log\left[\frac{P(\boldsymbol{\sigma}|m,x)}{P(\boldsymbol{\sigma}|m)}\right]  \right\rangle_{m,\boldsymbol{\sigma},x}\,.
\end{equation}
In this context, $I(\boldsymbol{\sigma};x|m)$ is PI carried by the pattern $\boldsymbol{\sigma}$ \emph{additional} to that in the morphogen signal $m$. It is zero if and only if $P(\boldsymbol{\sigma}|m,x)=P(\boldsymbol{\sigma}|m)$. It is easy to see that this condition is met  precisely when spatial interactions are zero. Then, $\boldsymbol{\sigma}(x)$ depends only on its local input, $m(x)$, and not directly on its location within the lattice. In other words, position, the morphogen signal, and the patterning genes form a Markov (dependency) chain, $x\rightarrow m\rightarrow\boldsymbol{\sigma}$. If, in contrast, patterning genes are spatially coupled, $\boldsymbol{\sigma}(x)$ additionally depends on the state of $\boldsymbol{\sigma}$ at neighboring lattice sites, which in turn respond also to their respective local morphogen signals. In that case, position, the morphogen signal, and the patterning genes do not form a Markov chain and DPI does not apply.

Our derivation and numerical results show that a spatially coupled network of genes downstream of the morphogen signal can extract PI in excess of that carried by the morphogen signal itself. Without spatial interactions, this is impossible in our setup, regardless of the complexity of local gene-gene interactions. Note that spatial interactions do not \emph{necessarily} lead to an increase in PI, but can do so if optimally chosen, in the regime where they permit spatial noise averaging.

\subsection{Long-range interactions enable Turing-like pattern formation, and make patterns robust to system size and morphogen signal variations}

Is there a role for spatial interactions beyond noise averaging? To address this question, we start by studying a seemingly unrelated problem of whether a Turing-like pattern generating mechanism also exists in our discrete setup. After establishing that is indeed the case, we proceed to show that in an appropriate limit Turing-like pattern generating capability also confers two biologically desired properties onto our system: the ability of the resulting patterns to automatically scale with the system size, and robustness to systematic perturbations of the morphogen signal, a special case of ``canalization'' that has been discussed in the biological literature \cite{lott2007,surkova2009,manu2009,jaeger2014}. 

\begin{figure*}
\includegraphics[width=12cm]{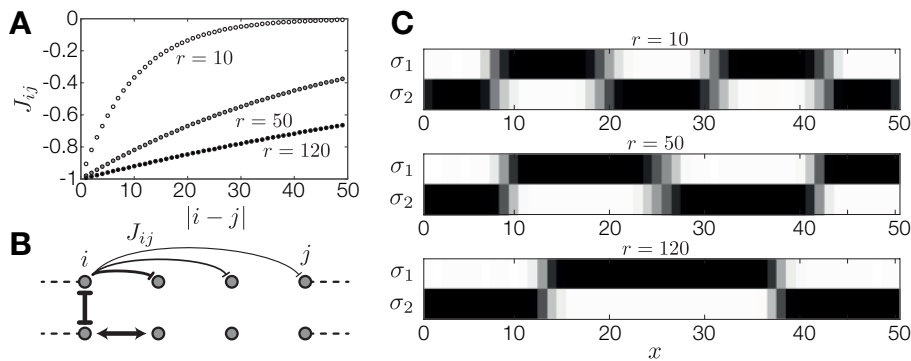}
\caption{
{\bf Long range spatial interactions can generate Turing-like patterns with an intrinsic length scale.} 
{\bf (A)} Spatial interaction strength of the first gene, $J_{11}(i,j) = \tilde J_{11} \exp(-(|i-j|-1)/r)$, as a function of distance $|i-j|$ between lattice sites for different interaction ranges, $r$. {\bf (B)} Schematic diagram of interactions in an extended model of patterning that permits long-range interactions. Interaction strength is indicated by the thickness of the arrows. Two genes in the model interact with a strong locally repressive interaction. The first gene has weak long range repressive spatial interactions; the second gene has strong nearest-neighbor positive interactions. {\bf (C)} Patterns generated by the two genes for three different interaction ranges, $r$. The expression state of the first gene is pinned to {\tt ON}  at the anterior boundary by a strong morphogen signal, which is zero anywhere else. The length scale of the resulting pattern depends on the interaction range, $r$, of the first gene. 
}
\label{fig4}
\end{figure*}

A distinguishing characteristic of Turing patterning is the emergence of patterns with an intrinsic length scale, which is set by the diffusion ranges of the two reacting species of the model. Until now, our model did not exhibit this property: spatial scale was either slaved to the external morphogen signal, as in the French Flag model, giving rise to boundary patterns; or the scale was one lattice spacing, as in the alternating pattern, where genes switched from {\tt ON} to {\tt OFF} at neighboring sites. 
Is it possible to generate patterns in which blocks of sites with a controllable intrinsic length scale alternate between {\tt ON} and {\tt OFF} states? To test if this behavior can arise in our model, we introduce long range spatial interactions. In particular, we assume that the spatial interaction strength between a gene $\alpha$ at lattice site $i$ and at lattice site $j$ decreases exponentially with distance: $J_{\alpha\alpha}(i,j) = \tilde J_{\alpha\alpha} \exp(-(|i-j|-1)/r)$, as in Fig.~\ref{fig4}A. Here, $\tilde J$ defines the amplitude of the interaction, which can also be negative. The parameter $r$ defines the interaction range, such that $r \rightarrow 0$ leads to nearest-neighbor interactions and $r \rightarrow \infty$ leads to a uniform, all-to-all coupling. 

To find a Turing-like pattern, we considered two mutually repressing genes, one of which interacts in a strong positive nearest-neighbor fashion, whereas the other has weak long range repression (Fig.~\ref{fig4}B). The resulting patterns for several interaction ranges $r$ are shown in Fig.~\ref{fig4}C. The system is clearly able to generate a blockwise alternating pattern with a length scale that  is controlled by the interaction range, $r$. We emphasize that the only morphogen signal  in this case is a strong positive bias at the anterior boundary, experienced by the first lattice site, provided to break the symmetry between the resulting pattern and its inverse. This explicitly demonstrates that the block length scale is not inherited from the shape of the morphogen signal, but is intrinsic to the interactions between the patterning genes. The same effect can be generated with a single gene, if we allow spatial interactions to change sign with distance (e.g., positive interaction at the nearest-neighbor range, and negative interaction with lattice sites at greater distance). 

Can we combine the ability of the Turing-like mechanism to generate intrinsic patterns with the network architecture that yields high-PI Counter patterns? The Turing mechanism has an attractive property which makes use of the morphogen signal only to break the symmetry between two possible intrinsically stable patters that are inverses of each other; in all other respects, the pattern is invariant, or robust, to changes in the morphogen signal magnitude or shape. The notion that external signals simply serve to select one of the few stable patterns (attractors) of the system, while the properties of the patterns are generated by intrinsic interactions, is known as ``canalization'' in the biological literature, but also has a very long history in neuroscience and statistical physics. Another feature of patterning that appears beneficial in a biological context is the ability of the system to translate small variations in the overall system size into proportional variations in the resulting gene expression pattern, i.e., to scale the pattern system size. If the system exhibits such ``scaling,'' gene expression features, for instance boundaries between {\tt ON} and {\tt OFF} states, will occur at constant fractional coordinates in the system, rather than at constant absolute coordinates. Here we test whether elements of Turing-like patterning can provide ``scaling'' and ``canalization'' properties to our Counter networks.

Before proceeding, we give an intuitive account as to why negative long range spatial interactions can be of benefit. A particular property of Counter patterns is their balanced use of {\tt ON} and {\tt OFF} expression states. Upon perturbations to the morphogen signal or shifts in pattern position because of system size changes, this balance will be broken. We are thus looking for a modification to the energy function, Eq~(\ref{hamiltonian}) (alternatively, Eq~(\ref{hamiltonianfull}) for multiple genes), that would penalize any deviation away from such balance. The simplest way to implement this is to add a term of the form $-\tilde{J}\left(\sum_{x=1}^N \sigma(x)\right)^2$ (or equivalent term for multiple genes). The term in parenthesis will be zero if the {\tt ON} and {\tt OFF} states are exactly balanced, and will be positive on any deviations in balance. When $\tilde J$ is chosen to be negative, such deviations are disfavored. If one analytically expands the square and takes into account that $\sigma\in[-1,1]$, one finds that the term corresponds to a global, all-to-all (long range) negative interaction between any pair of locations in the lattice. Crucially, such a term is an additive addition to an optimized Counter network: because the optimized Counter exhibits the {\tt ON} / {\tt OFF} balance already, the new term doesn't change the Counter state itself, but only makes states deviating from it less likely. Realistically, the strength and range of such interaction cannot be made infinite; the relevant question is therefore whether canalization and scaling can be obtained with a $\tilde J$ of finite magnitude and range.

\begin{figure*}
\includegraphics[width=15cm]{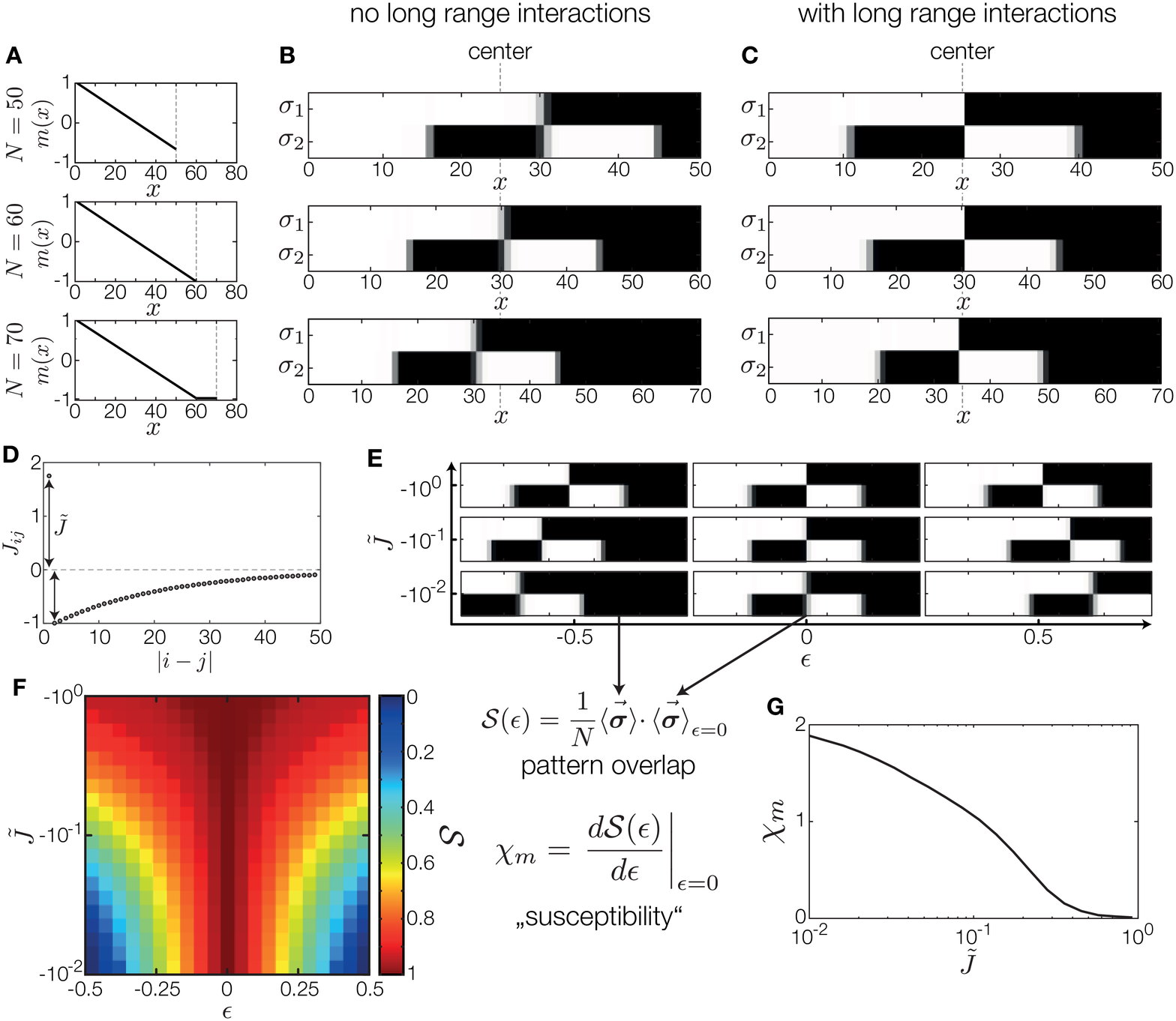}
\caption{
{\bf Long-range interactions can stabilize Counter patterns against variations in system size and morphogen signal.} {\bf (A)} Morphogen signal for three different system sizes, $N=50,60,70$. Posterior boundary of the system is  depicted as a dashed line. We consider how a two-gene Counter network, optimized for $N=60$, changes upon variations in system size. {\bf (B)} Resulting patterns for a Counter network for three system sizes depicted in (A). The pattern does not scale with the system size; instead, boundaries form at the same absolute location.  {\bf (C)} Resulting patterns for the same network as in (B) with additional negative long range interactions. Pattern shifts with system size are largely suppressed. {\bf (D)} Spatial interaction strength as a function of distance. Nearest neighbors are interacting positively, while sites further away are coupled negatively with exponentially decaying strength of maximal amplitude $\tilde J$. 
{\bf (E)} We perturb the morphogen signal with a uniform perturbation $\epsilon$, $\tilde{m}(x) = m(x) + \epsilon$. Example patterns generated by the optimal Counter two-gene network, at different strengths of long range interactions, $\tilde J$, and different morphogen signal perturbation magnitudes, $\epsilon$. At $\epsilon=0$, irrespective of $\tilde J$, the system generates the optimal pattern. When $|\epsilon|$ increases, the patterns shift (providing less PI) with weak $\tilde J$, but when $\tilde J$ is strong, the pattern is robust to such perturbations. {\bf (F)} To quantify the robustness to $\epsilon$ perturbations, we compute the overlap, $\mathcal{S}$, of the resulting pattern with the optimal Counter pattern. The overlap is shown as a function of $\epsilon$ and $\tilde J$; for strong negative $\tilde J$, the overlap is high irrespective of the perturbation strength, $\epsilon$. {\bf (G)} Susceptibility to small perturbations $\epsilon$ as a function of $\tilde J$ shows transition into a robust regime, $\chi_m\rightarrow 0$, as $\tilde J$ increases in magnitude.
}
\label{fig7}
\end{figure*}

We took the optimized two-gene Counter network and equipped it with long range negative spatial interactions, as shown in Fig.~\ref{fig7}D. We first tested whether this addition confers the scaling property to our Counter network. To this end, we consider a morphogen signal $m(x)$ which spans its dynamic range over a default system size of $N=60$ lattice sites. We model variations of the system size without scaling of the gradient by cutting the system and the gradient short by 10 lattice sites or extending it by 10 lattice sites at constant value of the morphogen signal in the posterior (Fig.~\ref{fig7}A). The corresponding gene expression patterns of the two-gene Counter network are depicted in Fig.~\ref{fig7}B. Without new long range spatial interactions, variations in system size do not result in the scaling of the pattern, as expected; instead, boundaries are fixed to their absolute positions. In contrast, when we add negative long-range interactions to each gene in our network, the pattern scales approximately, as shown in Fig.~\ref{fig7}C. Specifically, the central boundary is preserved with large precision, while the boundaries at $1/4$ and $3/4$ shift marginally. 

Next, we examine the robustness to systematic perturbations in the morphogen signal in detail. We introduce an additive offset $\epsilon$ to the morphogen signal, $\tilde{m}(x)= m(x) + \epsilon$, and ask about the resulting gene expression patterns with or without negative long-range spatial interactions. Note that additive perturbations in the morphogen signal map, in the thermodynamic model of gene regulation, to multiplicative perturbations in the morphogen concentration, $c(x)$. Such perturbations can be interpreted as variations in the morphogen dosage, and robustness to  such perturbations has been a focus of several experimental studies. Figure~\ref{fig7}E demonstrates that strong long-range spatial interactions indeed successfully make the pattern robust to (large) changes in $\epsilon$, while in the absence of such interactions the patterns experience large shifts with $\epsilon$. 

To quantify the stability of the pattern, we compute the overlap, $\mathcal{S}(\epsilon)=\frac{1}{N}\langle \vec{\boldsymbol{\sigma}}\rangle_{\epsilon=0} \cdot \langle \vec{\boldsymbol{\sigma}}\rangle_{\epsilon}$, as a function of the perturbation strength. An overlap of 1 means that under the perturbation the resulting pattern is identical---therefore fully robust---to the (Counter) pattern without perturbation; an overlap 0 means that on average half the expression states are inverted, while an overlap of -1 indicates that the perturbed pattern is the exact inverse of the unperturbed one. Mapping out the overlap as a function of perturbation strength, $\epsilon$, and the long range interaction magnitude, $\tilde J$, in Fig~\ref{fig7}E, we see that with sufficient, but still finite, $\tilde J$, the patterns can be made almost completely robust to large morphogen signal perturbations. A similar differential analysis in Fig~\ref{fig7}F shows how the susceptibility of the gene expression pattern to small morphogen signal perturbations vanishes as the strength of the long range interactions increases.

In sum, we have shown that ``canalization'' and ``scaling'' in our toy model system can be provided by the addition of long range negative spatial interactions. Curiously, these interactions can be added to a previously optimized Counter network without any need to change the optimized parameters, because they do not affect the energy or the identity of the Counter pattern ground state. As in the Turing model, these interactions stabilize the ground state. Unlike the Turing model, however, the Counter pattern itself is generated solely by a joint action of a French Flag-like morphogen signal readout and local gene-gene interactions. 

Should the long-range negative interactions that confer robustness also follow from an information optimization principle? Recalling our introductory remarks, robustness to, e.g., morphogen dosage $\epsilon$ could be measured directly as a low value of information $I(\epsilon;\vec{\boldsymbol{\sigma}})$, implying that changes in $\epsilon$ would not affect the expression patterns. The question is whether to gain robustness one needs to explicitly maximize PI while jointly minimizing $I(\epsilon;\vec{\boldsymbol{\sigma}})$, or whether it is sufficient to maximize PI only and get robustness as an automatic consequence. Due to the numerical complexity of the problem we did not perform such a large scale joint optimization here, but at non-zero intrinsic or extrinsic noise long-range interactions will stabilize the Counter pattern and thus maintain high PI beyond what could be achieved by local interactions alone. In the PI-maximization framework which we are proposing here, it thus seems that maximization of PI alone will also yield robustness, so long as the noise statistics under which the optimization is carried out contain the kinds of perturbations to which the system should be robust. As a conjecture, we suggest that, were we capable of carrying out large-scale optimization numerically, we would find optimal solutions with long-range negative couplings that yield canalization if our extrinsic noise also consisted of correlated additive fluctuations (that mimic the overall additive $\epsilon$ shifts in the morphogen signal considered above).

\section{Discussion}
We introduced a tractable toy model of patterning that extends Wolpert's French Flag model, in which several two-state genes respond to a morphogen signal based on a fixed set of thresholds. Our extension allows these genes to cross-regulate each other, to interact spatially (e.g., due to diffusion or cell-cell signaling), and to include the effects of both intrinsic noise (e.g., due to stochasticity in gene expression) and extrinsic noise (e.g., due to variability in the morphogen signal). A physics equivalent of our model is a set of coupled 1D Ising chains responding to inhomogeneous external field  and this link to statistical physics allows us to perform most of our computations exactly. In this well-defined setup we ask which patterns of gene expression encode the maximal amount of positional information; our enquiry is set in an optimization framework, where for each choice of noise magnitude and morphogen signal profile we look for the optimal pattern of gene-gene and spatial interactions, and examine the resulting spatial patterns of gene expression.

We find that with vanishing noise magnitude, the optimal gene expression pattern is the so-called Counter, where each of the $2^K$ binary patterns that can be realized by $K$ genes appears equally often along the spatial coordinate; this combinatorial code for position is, on information-theoretic grounds, the best achievable solution, independently of the patterning model. The Counter pattern can be realized if each patterning gene can be activated at a different threshold, and if the local gene-gene interactions can be adjusted; the optimal interactions are predominantly repressive. Increasing the noise quickly perturbs the Counter pattern, until the optimal pattern switches from Counter to French-Flag-like set of stripes, and finally, to a set of redundant genes that are all activated at the same threshold. Addition of short-range positive spatial interactions of intermediate strength makes the optimal patterns substantially more stable against increasing noise; qualitatively, this effect is analogous to noise averaging due to diffusive coupling in continuous systems. Strong short-range negative interactions generate an alternating pattern of gene expression which robustly separates odd from even rows of cells, a strategy that can yield one additional bit of PI when noise is low. 

Qualitatively new effects emerge if the spatial interactions can be long-ranged. We observe that our discrete, Ising-model-based model exhibits Turing-like patterns whose spatial scale is set by the range of the repressive interactions. Surprisingly, we find that the energy function, optimized to yield Counter patterns with high positional information, can be modified by the addition of strong long-range repressive interactions. This modification does not perturb the Counter pattern, but makes it robust to changes in the morphogen dosage and to changes in system size by producing patterns that approximately scale with system size. 

Taken together, our analysis allows us to identify elements---the basic building blocks---of positional information: adjustable thresholds as in the French Flag model to differentially drive gene activation; local repressive interactions to generate combinatorial codes for position; short-range positive spatial couplings to enable noise averaging and stabilize the patterns; and long-range negative spatial couplings to provide scaling and robustness via canalization. The optimal patterning system thus combines elements from both the French Flag model with the elements inherent to the Turing mechanism. Furthermore, these elements need not be identified and combined by hand, but emerge from a single information-theoretic optimization principle, and their contribution towards encoding of positional information can be individually quantified. 

The explicit purpose of this work has been to provide conceptual clarity and computational tractability rather than a detailed model of any particular patterning system. In order to achieve our goals, we had to sacrifice several crucial aspects of biological realism. First, real patterning does not happen at equilibrium, but is rather a driven dynamical process evolving from some initial state. Second, as the real patterning mechanisms are dynamic and might involve multiple timescales, noise mitigation mechanisms beyond spatial averaging, e.g., temporal averaging, should be available. Third, the assumption that expression states of patterning genes are binary might be poor. For example, in the gap gene system in \emph{Drosophila}, intermediate levels of expression are of crucial importance~\cite{dubuis2013}. On the other hand, regulatory circuits where individual genes have strong positive self-interactions and thus exhibit bistability could be well captured by our model. Fourth, the Ising framework assumes a particular (Boltzmann) distribution over expression states and thus leaves no degree of freedom to describe intrinsic stochasticity beyond its magnitude; in contrast, gene expression noise in real regulatory networks has a complicated relation to the mean expression~\cite{Tkacik:2008tp}. Last but not least, because our model is a statistical physics model at equilibrium, interactions between the genes are necessarily symmetric, which does not need to be the case in realistic gene regulatory networks. 

Despite these approximations, our model qualitatively recapitulates many aspects of the optimal patterning solutions that have been reported previously in more realistic setups, where the required computations are substantially more complicated~\cite{Tkacik_PRE1, Tkacik_PRE2, Tkacik_PRE3, Tkacik_PRE4, Tkacik_PRE5, erdmann2009}. This is in line with our previous observations that many mechanistic details that define the pattern-forming system $Q$ do not matter so long as the system has the ability to access patterns that achieve high positional information (which can be found by optimization). Pattern-forming processes can be complicated and can include spatial interactions between nearby cells or even long-range or global interactions. Ultimately, however, these systems generate patterns whose power is quantified by a local quantity, the positional information $I$, which is sufficient to summarize the limits to readout error. For optimal solutions to the patterning problem, this locality of positional code is crucial: there are many possible pattern-forming systems, but only in a restricted subset do we expect high information about global position to be available locally.


%
 \bibliographystyle{unsrt}
 \bibliography{./bibliography} 
%
 \end{document}